\documentclass[11pt,twocolumn,oneside]{article}
\usepackage[a4paper,top=3cm,bottom=3cm,left=1.7cm,right=1.7cm]{geometry}
\usepackage[babel]{csquotes}
\usepackage{graphicx}
\usepackage{subfigure}
\usepackage{amsmath,amssymb}
\usepackage[colorlinks,citecolor=blue,linkcolor=blue,urlcolor=blue]{hyperref}
\usepackage{mathrsfs}
\usepackage{authblk} 
\usepackage{abstract}
\usepackage{titlesec} 
\usepackage[T1]{fontenc}
\usepackage[margin=4pt, font=footnotesize, labelfont=bf, labelsep=period]{caption}	[2004/07/16]	

\newcommand{\Kmax}{{\cal K}^2_{\rm max}}
\newcommand{\alpla}{\alpha_{\rm max}}

\renewcommand\thesection{\Roman{section}}

\titleformat{\section}{\large\scshape\bfseries\centering}{\thesection.}{.7em}{}
\titleformat{\subsection}{\scshape\bfseries}{\thesubsection.}{.7em}{}

\def\be{\begin{equation}}
\def\ee{\end{equation}}

\title{\Large{\textbf{\textsc{On the Effective Metric of a Planck Star}}}}

\author[1,3]{Tommaso De Lorenzo\thanks{tommaso.de-lorenzo@cpt.univ-mrs.fr}
}
\author[2,3]{Costantino Pacilio\thanks{cpacilio@sissa.it}
}
\author[3]{Carlo Rovelli%\thanks{rovelli@cpt.univ-mrs.fr}
}
\author[3]{Simone Speziale%\thanks{simone.speziale@cpt.univ-mrs.fr}
}

\affil[1]{\textit{Universit\`a di Pisa, Dipartimento di Fisica ``Enrico Fermi'', Largo Bruno Pontecorvo 3, 56127 Pisa, Italy}}
\affil[2]{\textit{SISSA, Via Bonomea 265, 34136 Trieste, Italy}}
\affil[3]{\textit{Aix Marseille Universit\'e, CNRS, CPT, UMR 7332, 13288 Marseille, France \& Universit\'e de Toulon, CNRS, CPT, UMR 7332, 83957 La Garde, France.}\vspace{-2.5ex}}
\date{\small(Dated: \today)\vspace{-4ex}}

\begin{document}
\twocolumn[
\vspace{-7ex}\maketitle%
\begin{list}{}{\leftmargin=3em\rightmargin=\leftmargin}\item\relax
\small\textbf{\textsc{Abstract.}} 
Spacetime metrics describing `non-singular' black holes are commonly studied in the literature as effective modification to the Schwarzschild solution that mimic quantum gravity effects removing the central singularity. Here we point out that to be physically plausible, such metrics should also incorporate the 1-loop quantum corrections to the Newton potential and a non-trivial time delay between an observer at infinity and an observer in the regular center. We present a modification of the well-known Hayward metric that features these two properties. We discuss bounds on the maximal time delay imposed by conditions on the curvature,
and the consequences for the weak energy condition, in general violated by the large transversal pressures introduced by the time delay. 
\end{list}\par\vspace{3ex}%
]
\saythanks
\section*{Introduction}
Spacetime singularities are unavoidable in gravitational collapse, if classical general relativity is valid at all scales and the energy-momentum tensor of matter satisfies the classical energy conditions \cite{wald2010general, hawking1973large}. On the other hand, classical general relativity cannot be valid at all scales, because of quantum mechanics, and quantum effects such as Hawking radiation do violate these classical energy conditions. There is thus a certain expectation that near the center of a physical black hole quantum effects dominate, and prevent the formation of a singularity.
This scenario is also supported by a result of loop cosmology~\cite{Ashtekar:2006rx}: when matter reaches Planck density, quantum gravity generates pressure sufficient to counterbalance weight. For a black hole, this implies that matter's collapse can be stopped before the central singularity is formed, yielding to the formation of a central core, called a `Planck star' in \cite{rovelli:planckstars}. 
This expectation motivates the study of models of non-singular black holes, of which many examples exist in the literature (e.g. \cite{bardeen1968non, Frolov:1981mz, Roman:1983zza, Mazur:2001fv, dymnikova2002cosmological,hayward2006formation, Nicolini,Falls:2010he,ModestoNicolini,rovelli:planckstars,Bambi:2013ufa,Bambi:2013caa,frolov:BHclosed,Haggard:2014rza,Mersini-Houghton:2014yq,Frolov:2014wja}, and \cite{ansoldi2008spherical}  for a review). 

Most metrics in the literature, however, possess two characteristics which we find unphysical: firstly, a clock in the regular center is not delayed with respect to a clock at infinity; secondly, they do not reproduce the 1-loop quantum corrections computed in~\cite{BjerrumBohr:2002kt} treating quantum general relativity as an effective field theory. 
In this short note, we show how it is possible to write effective line elements incorporating these two effects. In particular, we propose an explicit and simple modification of the Hayward metric that achieves the desired result, and discuss limitations on the maximal time delay allowed by the condition that the curvature remains sub-Planckian everywhere. Finally, we expose how our modification introduces a violation of the weak energy condition in a small region around the internal horizon, and write down conditions for its avoidance. We restrict our considerations to spherically symmetric and static metrics. Applications of our proposed modified Hayward metric to dynamical scenarios is left to future work. We use natural units $c=G=\hbar=1$ throughout the paper.

%%%%%%%%%%%%%%%

\section{Hayward Metric}\label{sec:hayward}
Consider the spacetime of a spherically symmetric 
object described by a static metric of the type%
\be \label{eq:dsa0=0}
ds^2 = - F(r) dt^2 + \frac{1}{F(r)} dr^2 + r^2 d\Omega^2
\ee
with
\be\label{eq:F(r)M(r)}
F(r) = 1- \frac{2 M(r)}{r} \;.
\ee	
Here $M(r)$ tends to a constant value $m$ for $r\mapsto\infty$, so to recover the Schwarzschild solution as a large distance approximation, but it is such that the spacetime is nowhere singular. Various choices for $M(r)$ have appeared in the literature (e.g. \cite{bardeen1968non,dymnikova1992vacuum,Mazur:2001fv,hayward2006formation}). In the following we work with the metric originally presented by Hayward in~\cite{hayward2006formation} and recently reconsidered, focusing on quantum gravitational phenomenology, in~\cite{frolov:BHclosed,rovelli:planckstars}. 
For the Hayward metric,
\be\label{eq:M(r)}
M(r) = \frac{m\,r^3}{r^3 + 2 m L^2},
\ee
where $L$ is a parameter with dimensions of a length. The logic here is that such a metric could arise from a low-energy limit of quantum gravity, as a solution to Einstein's equations modified by an additional right-hand side coming from the fundamental quantum theory. In this context, $L$ is a free parameter that is natural to assume of the order of the Planck length. 
Specifically, the right-hand side as the form of a diagonal energy-momentum tensor with
\begin{align}
& \rho=\frac1{2\pi} \frac{3L^2m^2}{(2L^2 m + r^3)^2} = -p_r, \\
& p_t=-\frac1{\pi} \frac{3L^2m^2 (L^2m-r^3)}{(2L^2 m + r^3)^3}.
\end{align}
These expressions are compatible with the weak energy condition everywhere, but violate the strong energy condition $\rho+p_r +2p_t\geq 0$ for $r^3\leq L^2m$. 
Indeed, this can be put in evidence noticing that near the origin the metric behaves like a de Sitter spacetime, 
\be\label{eq:desitter}
F(r) = 1- \frac{r^2}{L^2} +o(r^3),
\ee
with the effective cosmological constant $\Lambda=3/L^2$ introducing a repulsive force and thus violating the strong energy condition. It is this repulsive force that is responsible for the avoidance of the singularity, in this logic introduced by quantum gravity effects, as in the example of the bounce in loop cosmology.

\begin{figure}[t]
\centering
\includegraphics[width=0.45\textwidth]{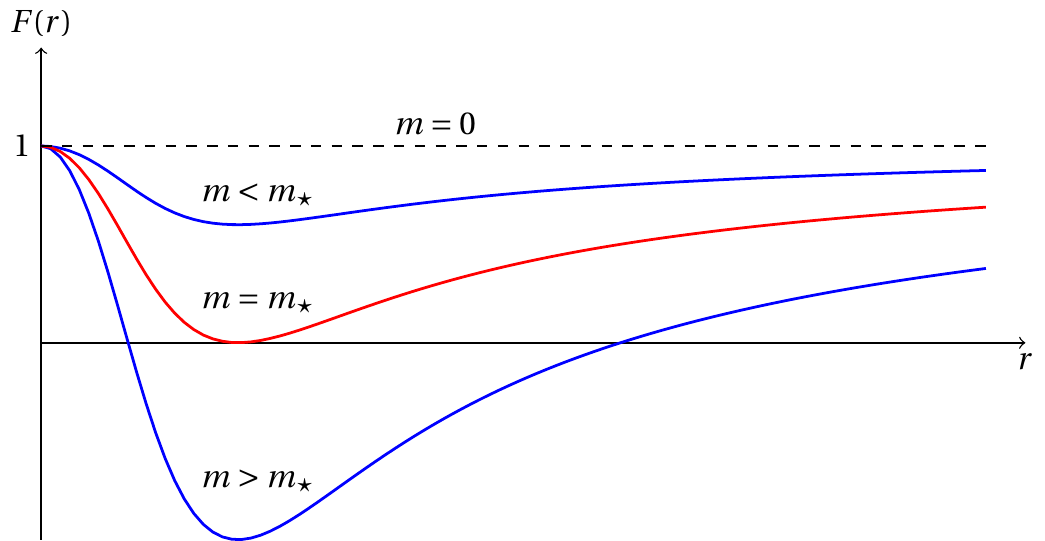}
\caption{Redshift factor $-g_{00}$ of the Hayward metric as a function of the radius for different values of the mass $m$.}
\label{fig:F}
\end{figure}

\begin{figure}[t]
\centering
\includegraphics[width=0.45\textwidth]{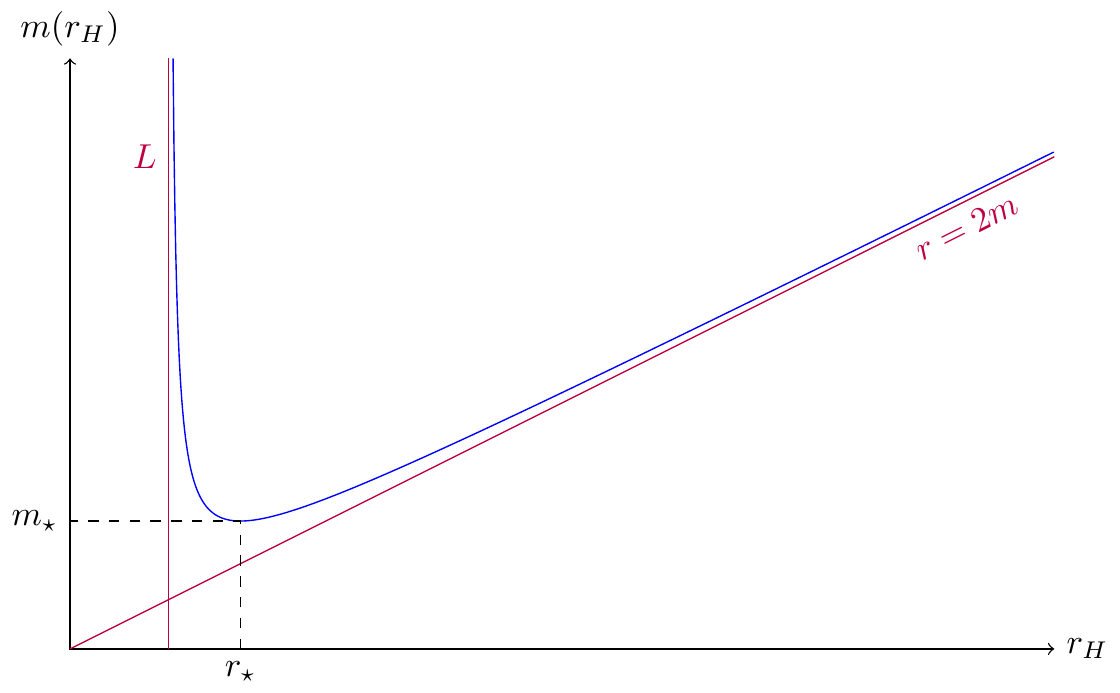}
\caption{The relation between the location of the horizons and the mass, at fixed $L$. Asymptotic values are reached for $m\gg L$.}
\label{fig:horizons}
\end{figure}

The spacetime under study possesses Killing horizons when $F(r)=0$. With $M(r)$ given by \eqref{eq:M(r)}, the existence of solutions is controlled by the parameters $m$ and $L$. There are two Killing horizons for $m > m_\star = \frac{3\sqrt{3}}{4} L$, merging into one at the critical value $m=m_\star$, see Fig.~\ref{fig:F}.
The position of the horizons is given implicitly by 
\be
m(r_H) = \frac{r_H^3}{2(r_H^2 - L^2)},
\ee
and for $m \gg L$ the inner and the outer horizons approach respectively $r_- \simeq L$ and $r_+ \simeq 2m$. See Figure~\ref{fig:horizons}.
Therefore, for values of the mass greater than the critical value $m_\star$, the metric features an outer horizon and can be taken as a description of a non-singular black hole.

\begin{figure}[t]
\centering
\includegraphics[width=0.35\textwidth]{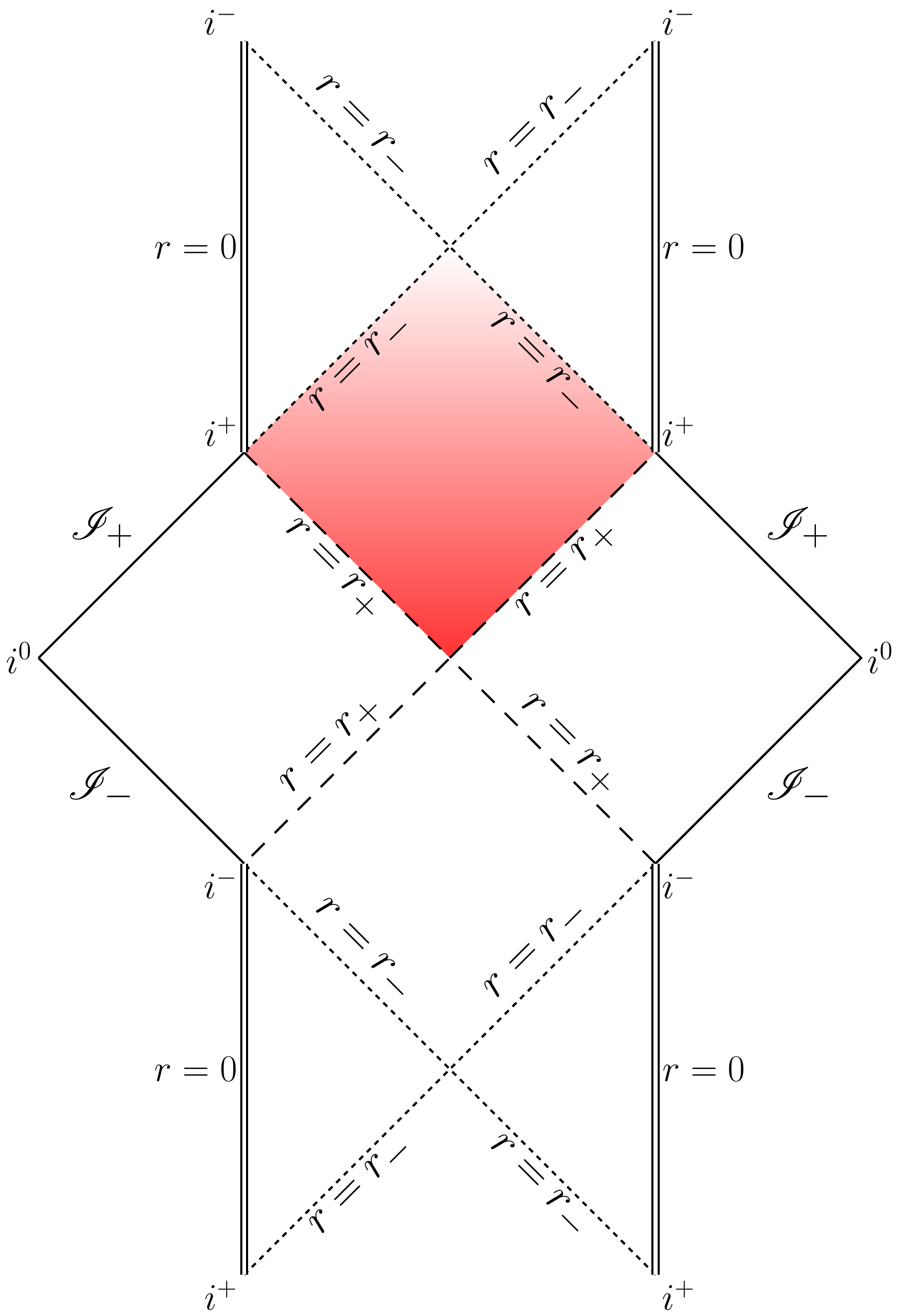}
\caption{Penrose diagram for the static line element in Eq.~\eqref{eq:dsa0=0}.}
\label{fig:penrose1}
\end{figure}

Since the metric violates only the strong energy condition, it is interesting to recall how the original Penrose's singularity theorem, based upon the weak energy condition, is also avoided. The causal structure of this metric, represented in the Penrose diagram of Figure~\ref{fig:penrose1}, is analogous to that of the Reissner-Nordstr\"om spacetime, with the difference that the time-like surface $r=0$ is not singular anymore. The surface $r=r_-$ is a Cauchy horizon which means that the spacetime is not globally hyperbolic, hence the Penrose's singularity theorem does not apply and the metric can be regular in spite of satisfying the dominant energy condition. 
The lack of global hyperbolicity can also be understood in terms of the topology change from a compact to a non-compact inner region, an argument which had already been shown to lead to avoiding the singularity theorems~\cite{Borde:1996df}.

\begin{figure}
\centering
\includegraphics[width=0.45\textwidth]{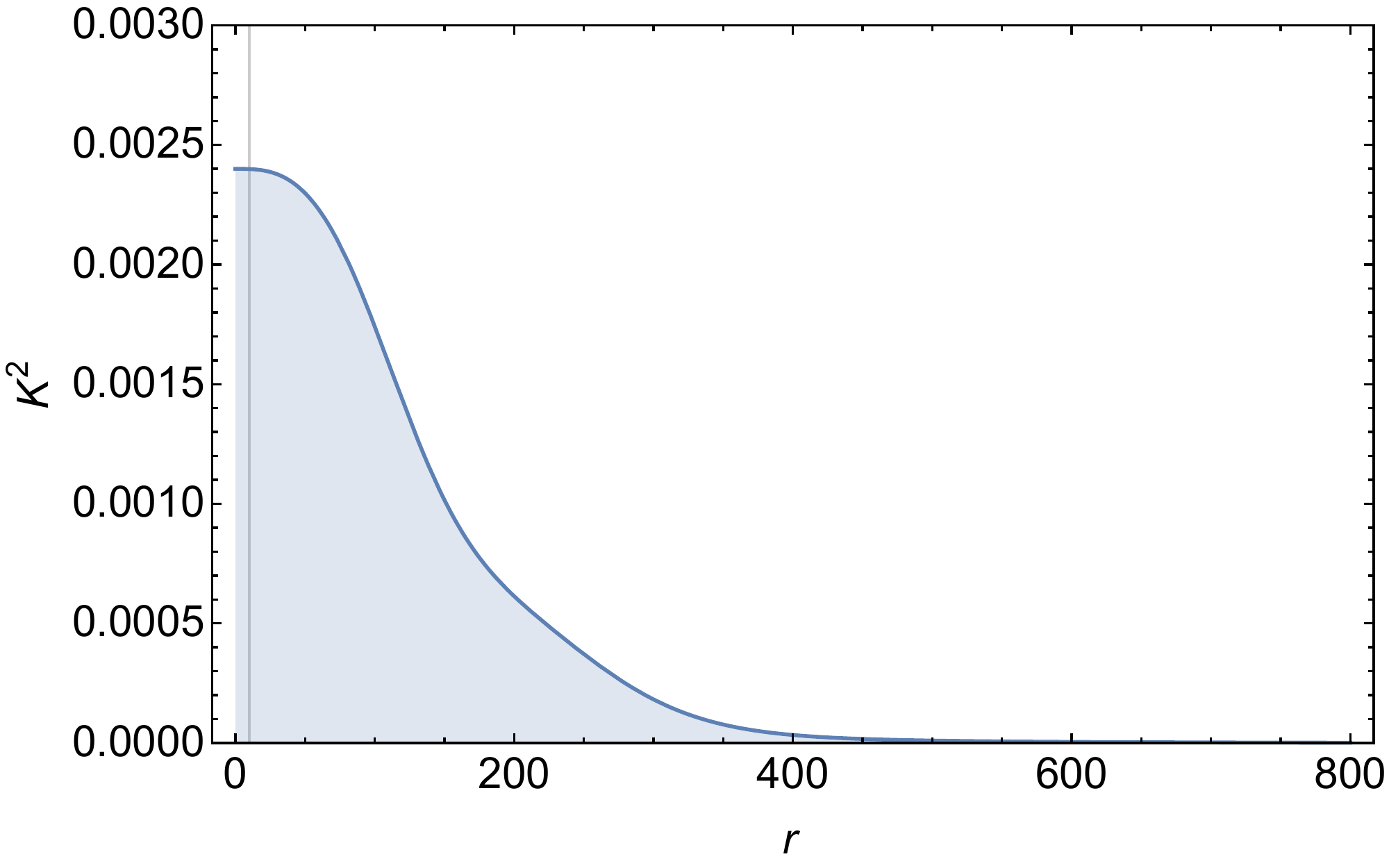}
\caption{Kretschmann scalar curvature as a function of the radius for the Hayward metric; here $m=10^5$ and $L=10$ (Planck units). The vertical line represents the inner horizon $r_-$.}
\label{fig:hayR}
\end{figure}

The violation of classical energy conditions is a natural consequence of the fact that the singularity-free metric is supposed to include some quantum gravity effects. On the other hand, restrictions on the metric come from the requirement that its curvature has to be sub-Planckian everywhere, as discussed for instance in \cite{Frolov:1989pf}, so that details of quantum gravity (such as the spacetime foam proposed by loop quantum gravity \cite{CarloBook}) do not matter, and the effective description is meaningful.
As a criterium for maximal curvature, we consider the Kretschmann scalar
${\cal K}^2 = R_{\mu\nu\rho\sigma}R^{\mu\nu\rho\sigma}$,  where $R_{\mu\nu\rho\sigma}$ is the Riemann tensor.
Its value as a function of $r$ for the Hayward metric is shown in Fig.~\ref{fig:hayR}. It smoothly decreases from a maximum in the origin to zero. Hence, the curvature will be sub-Planckian everywhere provided it is so near the origin. Taylor-expanding the exact expression we obtain
\be
{\cal K}^2 = \frac{24}{L^4} \left(1 - 2 \frac{r^3}{m L^2}\right)+ o(r^5)\;.
\ee
Therefore, requiring ${\cal K}^2\leq 1$ imposes a restriction on $L$, and on $L$ alone, $m$ controlling only the slope of the curve and not its maximal value. Specifically, there is a lower bound $L \gtrsim 3$ in Planck units. 

\begin{figure}[t]
\centering
\includegraphics[width=0.3\textwidth]{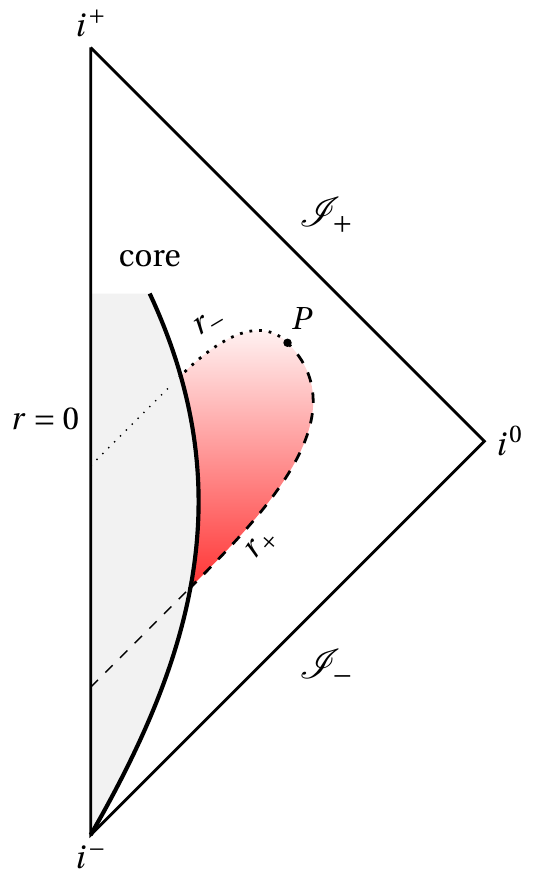}
\caption{Penrose diagram of a collapsing and evaporating non-singular black hole. The thick line is the external boundary of the star, while the trapping region is shaded, bounded by the two trapping horizons: the external evaporating one, and the internal expanding one.}
\label{fig:penrose2}
\end{figure}

The static Hayward metric has been suggested as an effective metric to describe a certain stage of the life of a black hole, occurring between the initial collapse and the moment when the evaporation becomes important \cite{hayward2006formation,frolov:BHclosed,Frolov:2014wja}, with related applications to the information-loss puzzle (see e.g. \cite{Ashtekar:2005cj,Bianchi:2014bma,Perez:2014lq}). As energy is radiated away in the form of Hawking radiation, the mass $m$ decreases until the critical value $m_\star$ and a regular gravitating object without horizons remains. In the Planck star scenario \cite{rovelli:planckstars}, the initially collapsing matter bounces out. The time dependence of the internal horizon, and hence the radius of the `explosion' event $P$ depends on the quantum gravitational dynamics of the Planck star. Because of the expected huge time dilation inside the gravitational potential well of the star, the bounce is seen in extreme slow motion from the outside, appearing as a nearly stationary black hole. Furthermore, the core (the Planck star) retains memory of the initial collapsed mass $m$ and the final exploding objects is much larger than Planckian. See \cite{Haggard:2014rza,Barrau:2014hda,Barrau:2014yka} for developments and applications to observations of this model. 
The process is illustrated by the conformal diagram of Figure~\ref{fig:penrose2}. 
While the static metric plays an important role in this compelling picture, it has two shortcomings that we want now to point out.

\subsection*{Two Shortcomings}

Gravitational time dilatation slows down clocks in a gravitational potential well, compared to clocks in an asymptotically flat region.  A clock kept in the centre of a dust cloud, for example, shows an elapsed time shorter than that of a clock at infinity, when the two clocks are moved together and compared. Since the Hayward metric is regular at the origin, we can imagine a clock sitting at the centre of the collapsed object.  
The time measured by this clock is easy to compute: during the static phase $F(r=0)=1$, Eq.~\eqref{eq:dsa0=0} shows that this is equal to the coordinate time $t$. The same is true for a clock at infinity. Therefore a clock at the centre of the star suffers no time delay with respect to a clock at infinity, and a signal sent from infinity with a given frequency will be received at the center with the same frequency. 
This is a physically unmotivated restriction, and directly related to the fact that $F(0)=1$. Notice that this feature is shared by most models in the literature, one exception being the gravastar of \cite{Mazur:2001fv}.

There is a second limitation of \eqref{eq:M(r)} that we would like to point out. While this effective metric is mainly motivated by including strong quantum gravity effects, it should not neglect the inclusion of those weak, but solid, quantum gravity effects that are commonly agreed upon, such as the 1-loop quantum corrections to the Newton potential obtained using effective field theory \cite{BjerrumBohr:2002kt,Donoghue:2012zc}. The latter read, reintroducing the Planck length,
\begin{equation}
\label{eq:donoghue}
\Phi(r)=-\frac{m}{r}\left(1 + \beta \,\frac{\ell^2_{\text{planck}}}{r^2}\right)+o(r^{-4})\,,
\end{equation}
where $\beta$ is a numerical constant of order 1.
By a standard derivation of the Newton potential from the $g_{00}$ component of the Schwarzschild metric, 
\be
\Phi(r) = -\frac{1}{2}\big(1+g_{00}(r)\big),
\ee
the 1-loop effect is immediately related to the metric. 
Such an effect cannot be described by Hayward metric (\ref{eq:M(r)}), whose large scale behaviour is
\be
g_{00} = -1+\frac{2m}r -\frac{4L^2 m^2}{r^4}+o(r^{-5})
\ee
and lacks the appropriate $r^{-3}$ term.\footnote{It is worth mentioning that the metric proposed by Bardeen in~\cite{bardeen1968non} does reproduce the required behaviour of the newtonian potential. On the other hand, as well as all the line elements proposed, it suffers for the time delay problem.} 

We now introduce a minimal generalisation of the Hayward metric that allows to fit in these two requests.

%%%%%%%%%%%%%%%%%%

\section{Modified Hayward Metric} \label{sec:modhayward}
The most general spherically symmetric, static metric can be parametrized adding an arbitrary function $G(r)$ to the $00$ component of the metric,
\be\label{eq:linelement}
ds^2 = - G(r)F(r) dt^2 + \frac{1}{F(r)} dr^2 + r^2 d\Omega^2.
\ee
In the following we take the same $F(r)$ as Hayward, and use $G(r)$ to introduce the desired modifications.
From the previous discussion, the physical requirements we wish to impose on $G(r)$ are:
\begin{description}
\item[$(i)$] preserve the Schwarzschild behaviour at large $r$;
\item[$(ii)$] include the 1-loop quantum corrections \eqref{eq:donoghue};
\item[$(iii)$] allow for a finite time dilation between the center and infinity.
\end{description}
A time delay between the center and infinity can be seen from
$(\delta t_{\infty} - \delta t_0)/\delta t_\infty = 1- \sqrt{|g_{00}(r=0)|} \in [0,1)$, and we parametrise
\be\label{def-eps}
- g_{00}(r=0) = 1 -\alpha, \qquad \alpha\in[0,1).
\ee
The larger $\alpha$, the greater the time delay.

The first two conditions are satisfied by 
\be\label{eq:conditiondon}
\lim_{r\to \infty} G(r) = 1- \beta\; \frac{m\, \ell^2_{\text{planck}}}{r^3}\;
\ee
where $\beta$ is the same of Eq.~\eqref{eq:donoghue}. In the remaining of this paper, we consider it fixed once and for all.\footnote{In the numerical plots, we use the value $\beta=41/10\pi$.} At $r=0$, condition $(iii)$ gives
\be\label{eq:epsilon}
G(0) = 1-\alpha.
\ee
An additional useful restriction -- albeit not mandatory -- is to demand that $\emph{(iv)}$ near the center, the equation of state of the derived energy momentum tensor is still de Sitter. Since
\begin{align}
g_{00}(r) = &- G(0)\left(1-\frac{r^2}{L^2}\right) - G'(0) r -\\ \nonumber
& - \frac{G''(0)}{2}r^2 + o(r^3), 
\end{align}
matching the de Sitter behaviour (cf. \eqref{eq:desitter}) gives
\be\label{eq:G(0)}
G'(0)=G''(0) = 0\;.
\ee
$G(0)$ can then be absorbed rescaling $t$, introducing in this way the desired time delay.
Conditions $(i)$ and $(iv)$, and their associated expressions \eqref{eq:conditiondon} and \eqref{eq:G(0)}, suggest to look for solutions as rational functions of $r^3$. Taking the simplest case, and using $(ii)$ and $(iii)$ to fix the coefficients, we find
\be\label{eq:G}
G(r) = 1 - \frac{\beta m\,\alpha}{\alpha\, r^3+\beta m}.
\ee
This example shows how it is possible to improve the metric proposed by Hayward to take into account the 1-loop quantum corrections and a time delay in the central core.

\bigskip

\subsection*{Bounds} 

\begin{figure}[t]
\centering
\includegraphics[width=0.45\textwidth]{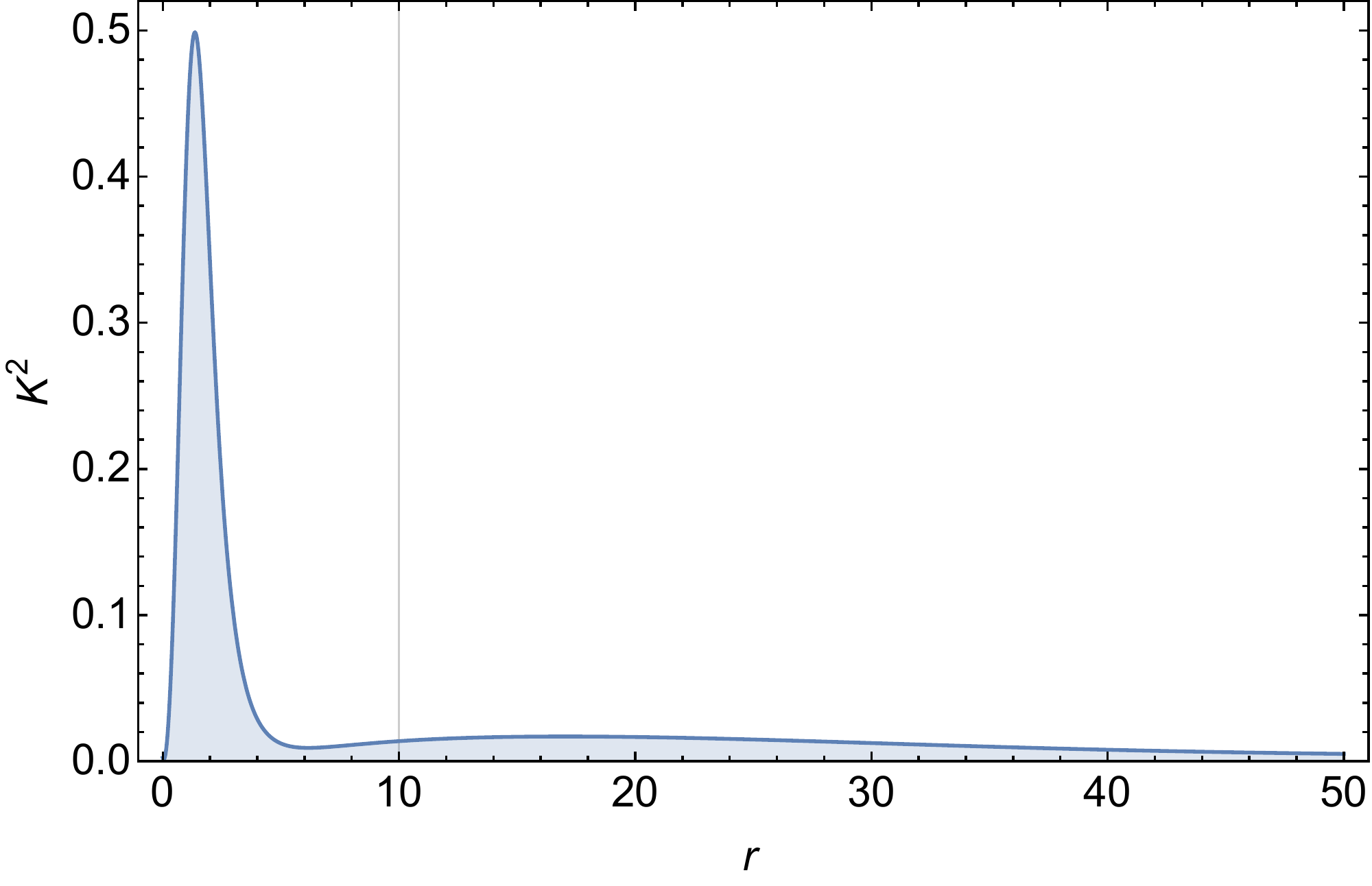}
\caption{Kretschmann scalar curvature as a function of the radial coordinate $r$ for the corrected Hayward metric. A peak features inside the inner horizon (marked by the vertical line), whose value increases beyond the Planck scale if the required time delay is too large. In this plot $1-\alpha = 7\times 10^{-5}$, $m=10^5 m_{\rm Planck}$, $L=10 \ell_{\rm Planck}$ and the peak is shown at $1/(2{\ell_{\rm Planck}^4})$. Notice also that the modification introduces a decreasing of the curvature near the inner horizon, and before the peak.}
\label{fig:TransPlanck}
\end{figure}

Next, we check what restrictions apply to the new metric.
In particular, we will see that it is not possible to arbitrarily increase the time delay between the center and infinity: an upper bound comes from the requirement that the curvature is sub-Planckian everywhere. 
The Kretschmann invariant associated with (\ref{eq:G}) has a rather long expression which prevents a purely analytic study of its properties. However, numerical investigations clearly show that for $\alpha>0$ a peak in curvature develops inside the inner horizon, see Fig.~\ref{fig:TransPlanck}. The value of the peak can become arbitrarily large as $\alpha$ approaches 1, and exceeds the Planckian value $1/\ell_{\rm Planck}^4$ at a value that depends on $m$ and $L$.
Imposing that the maximal value of the curvature stays always below the Planck scale thus introduces an upper bound on $\alpha$.

\begin{figure}
\centering
\includegraphics[width=0.45\textwidth]{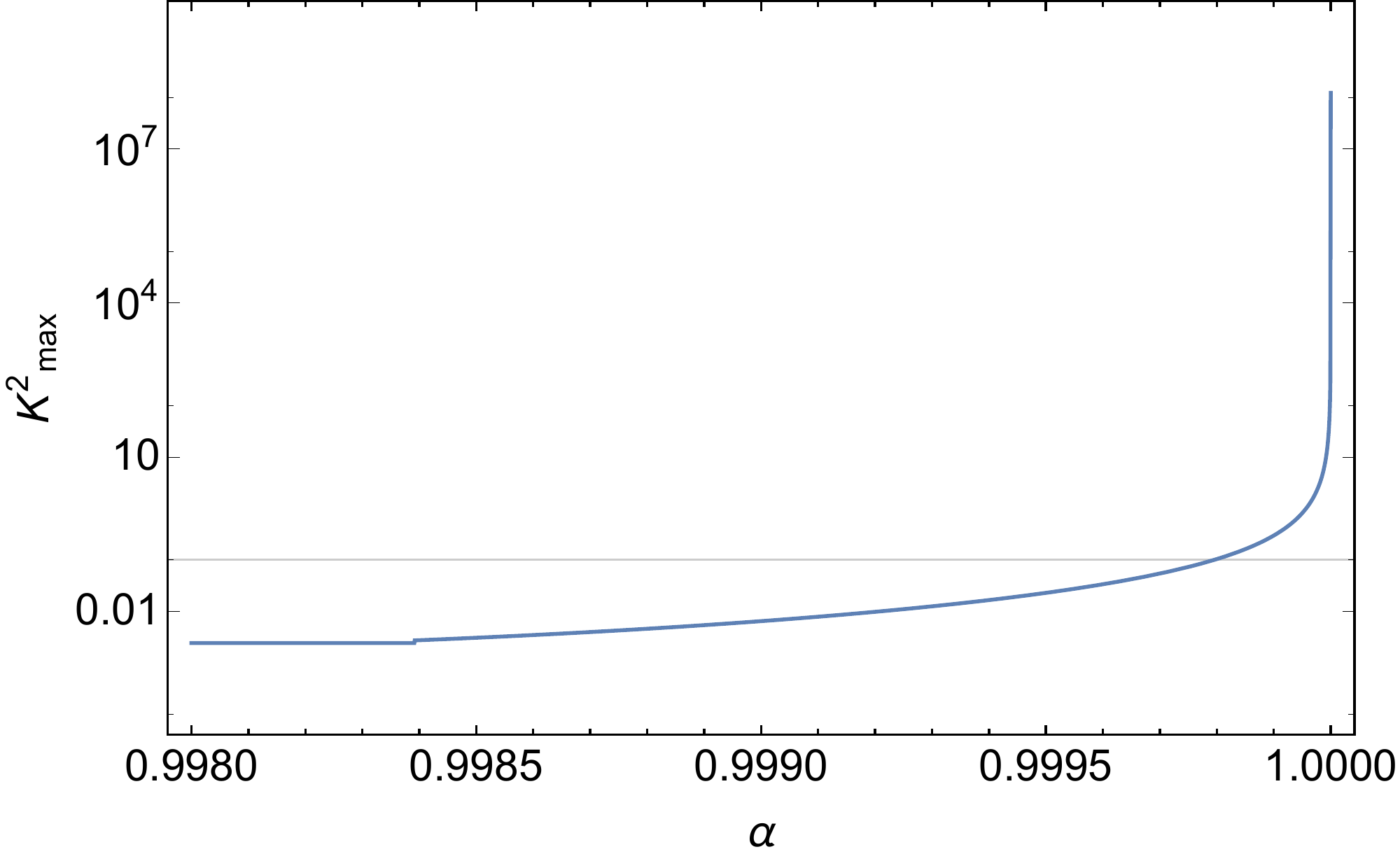}
\caption{Maximum value of the Kretschmann scalar as a function of $\alpha$. Here again $m=10^5 m_{\rm Planck}$ and $L=10 \ell_{\rm Planck}$, but the same behaviour is reproduced for all values of $L$ and $m>m_\star$.}
\label{fig:Kmax}
\end{figure}

Indeed, let us call $\Kmax$ the maximum value of the Kretschmann scalar curvature; in general it will be a function of the three free parameters of the model, i.e. $m$, $L$ and $\alpha$. 
The numerical analysis, see Fig.~\ref{fig:Kmax}, shows a monotonically increasing behaviour in $\alpha$ for $\Kmax$, namely
\be
\frac{\partial \Kmax}{\partial\alpha} > 0, \qquad \forall m>m_\star.
\ee
Therefore, we can impose a bound on $\alpha$ by requiring the maximum curvature to be smaller than unity, say $0.1$. Let us call $\alpha_\text{max}$ the bound value, namely
\be
\Kmax(\alpla,m,L) = 0.1	\;.
\ee
For instance $m=10^5$ and $L=10$ give $1-\alpla\sim 2\times 10^{-4}$, corresponding to a time delay of 
$98\%$. We thus see that the bound is not very stringent, and rather large time delays can be introduced without violating the condition that the metric can be taken as an effective description throughout spacetime.

\begin{figure}
\centering
\includegraphics[width=0.45\textwidth]{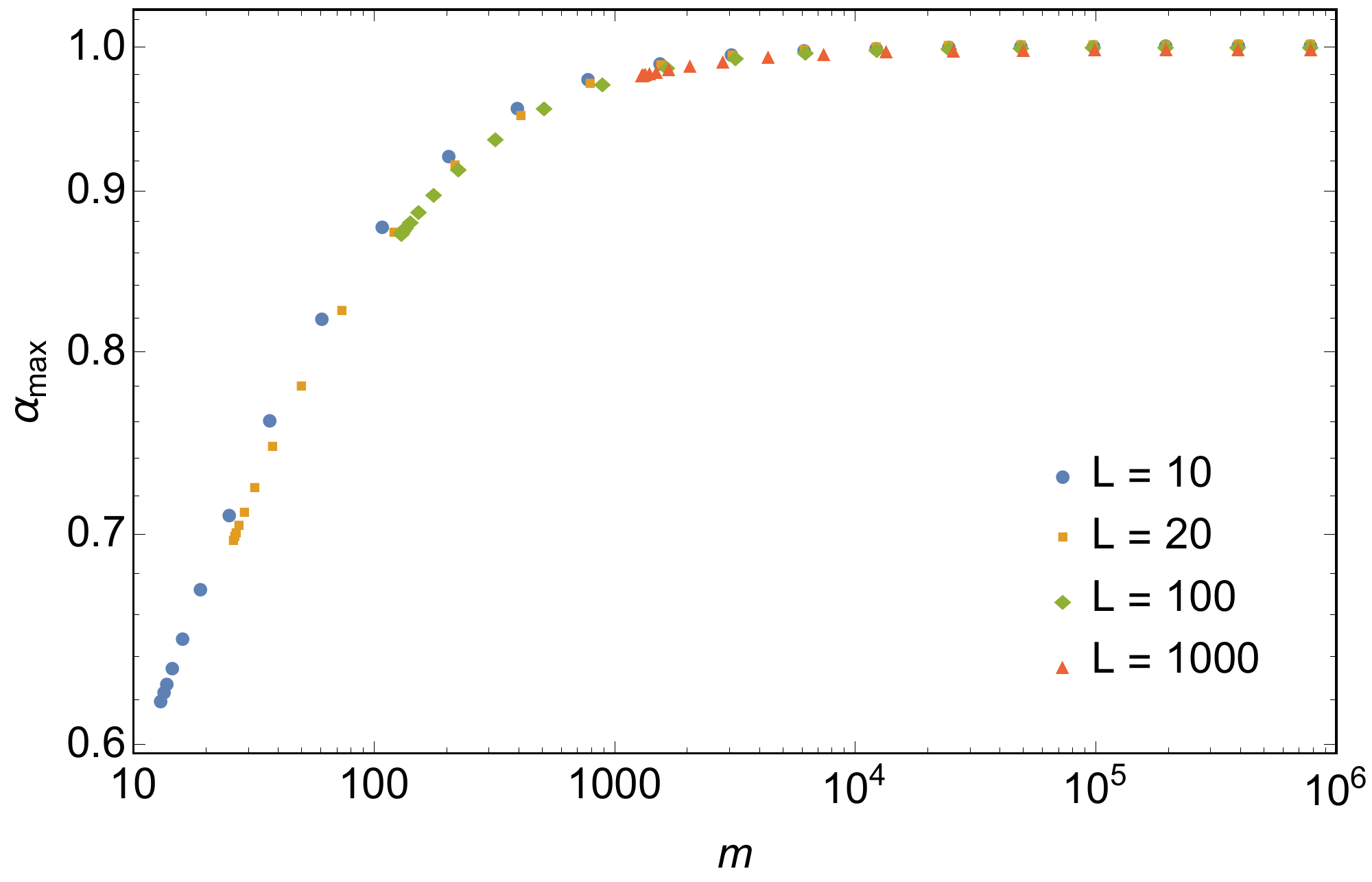}
\caption{$\alpha_\text{max}$ as a function of the mass $m$ for different values of $L$ (Planck units).}
\label{fig:alphLmult}
\end{figure}

The dependence of the bound on the two other parameters of the model is reported in Fig.~\ref{fig:alphLmult}. The plots show that $\alpla$ increases with $m$, and (slightly) decreases with $L$. In physical terms, a larger (non-singular) black hole allows a greater time delay, whereas stronger quantum gravity effects reduce it.

\begin{figure}[t]
\centering
\includegraphics[width=0.45\textwidth]{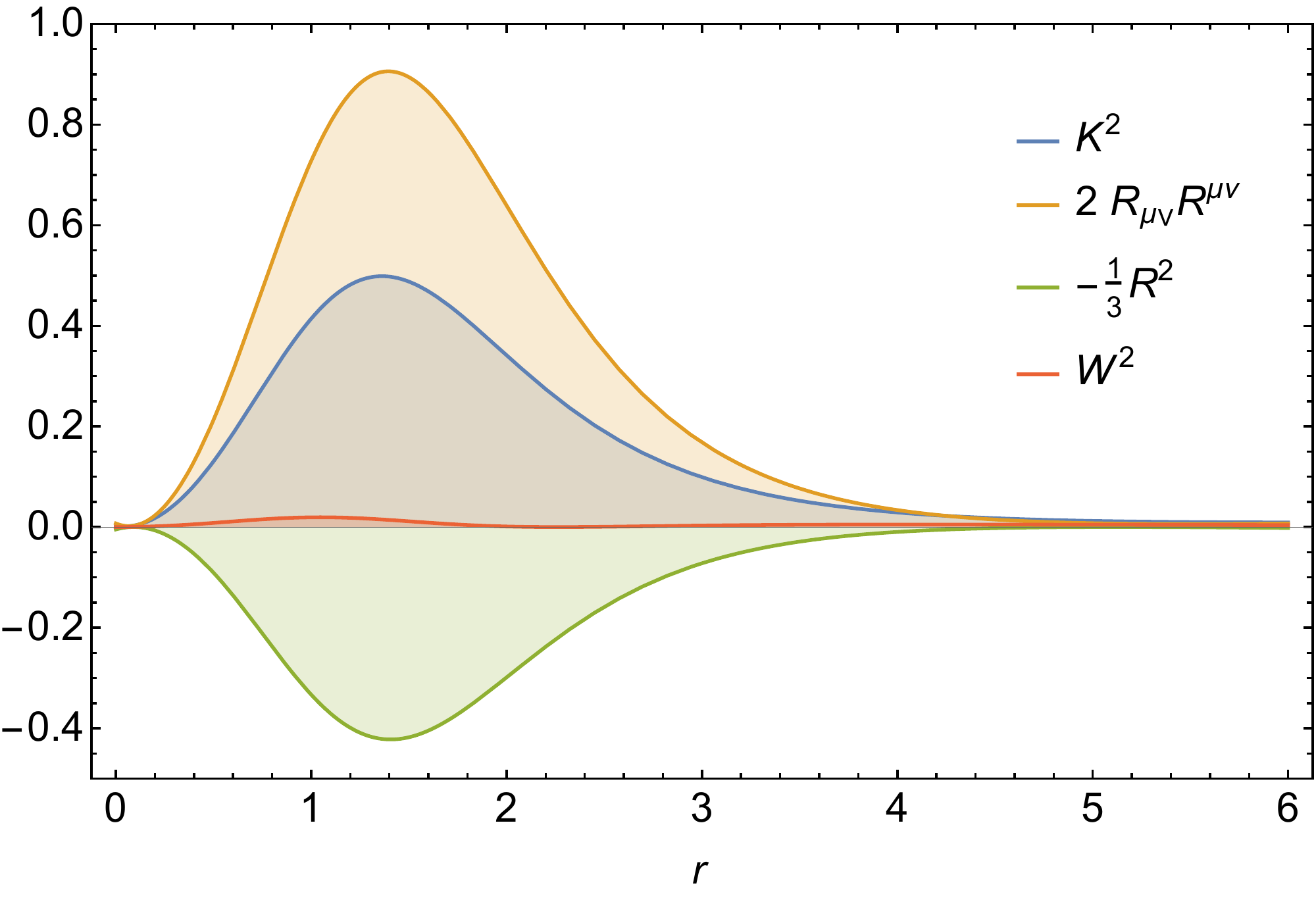}
\caption{Comparison between the different factor contributing to the Kretschmann scalar $\mathcal{K}^2$. The Weyl's tensor contribution is always small. Same parametrical values as in the previous figure.}
\label{fig:Weyl}
\end{figure}

The origin of the peak, and thus of the bound, is to be found in the behaviour of the effective energy-momentum tensor for which the modified Hayward metric is a solution. Indeed,  in Fig.~\ref{fig:Weyl} we plot separately the various Weyl and Ricci contributions to the Kretschmann scalar, showing the Weyl contribution is always small, and the peak comes from the Ricci tensor uniquely.

\begin{figure}
\centering
\includegraphics[width=0.45\textwidth]{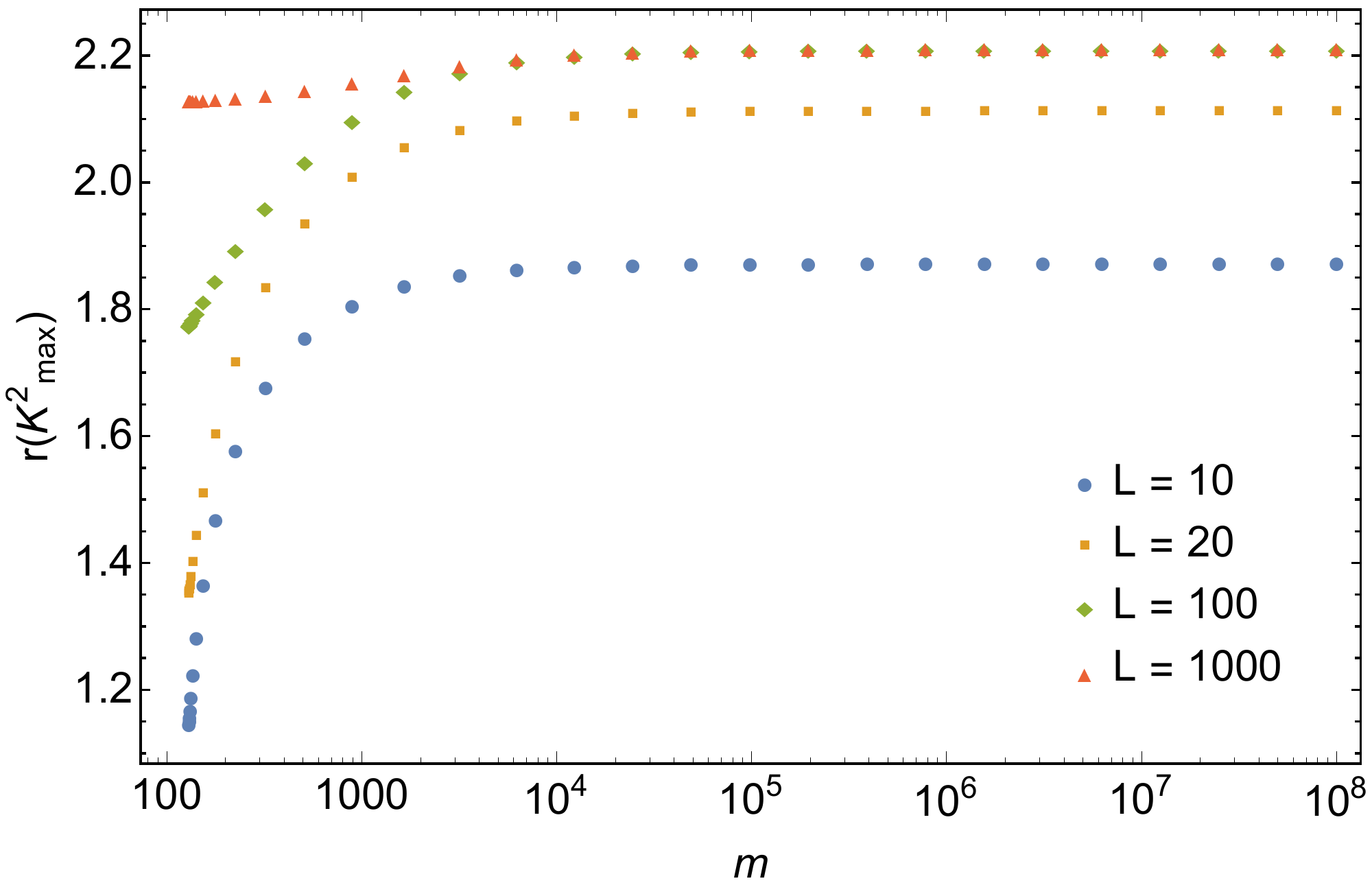}
\caption{The position of the peak, at its maximal allowed value $0.1$, as a function of $m$, for various values of  $L$. It reaches asymptotic values in both parameters, and it is always well inside the inner horizon.}
\label{fig:rKMax}
\end{figure}

Finally, the position of the maximal peak at $\alpla$ as a function of $m$ and $L$ is reported in Fig.~\ref{fig:rKMax}.

\section{Energy Conditions}
\begin{figure}[t]
\centering
\includegraphics[width=0.45\textwidth]{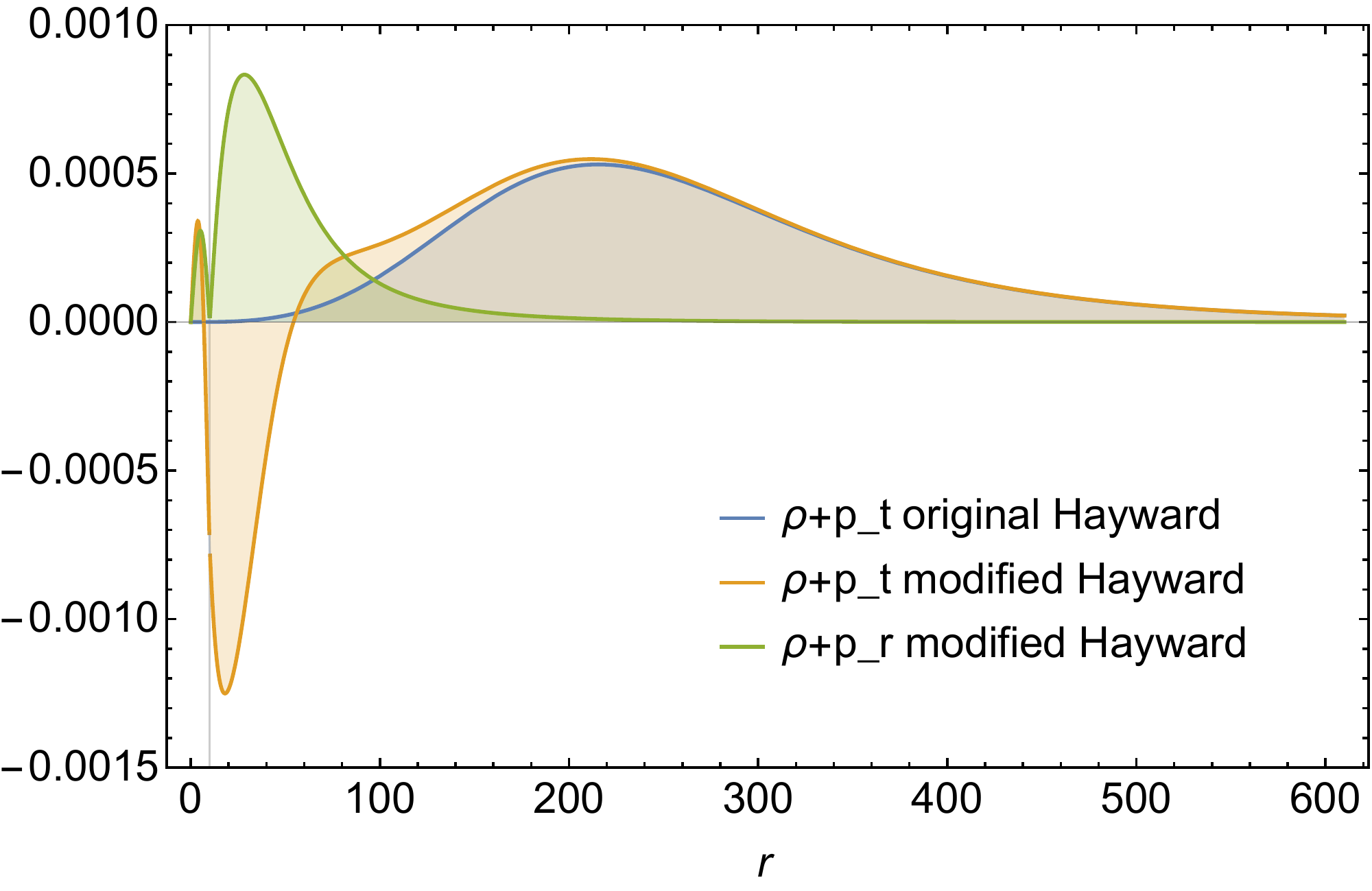}
\caption{Pressure profiles for both the original Hayward metric and its modification. The radial component of the weak energy condition is zero everywhere in Hawyard's case, and shown in green for the modified case. The transversal contribution, in blue for the original Hayward metric and in orange for the modified one. The latter shows the violation of the condition. 
The violation is confined to a small region inside the inner horizon, marked by the vertical line in the plot. Here $m=10^5 m_{\rm Planck}$, $L=10 \ell_{\rm Planck}$, and $1-\alpha = 0.99$.}
\label{fig:energy_violation}
\end{figure}

\begin{figure}
\centering
\includegraphics[width=0.45\textwidth]{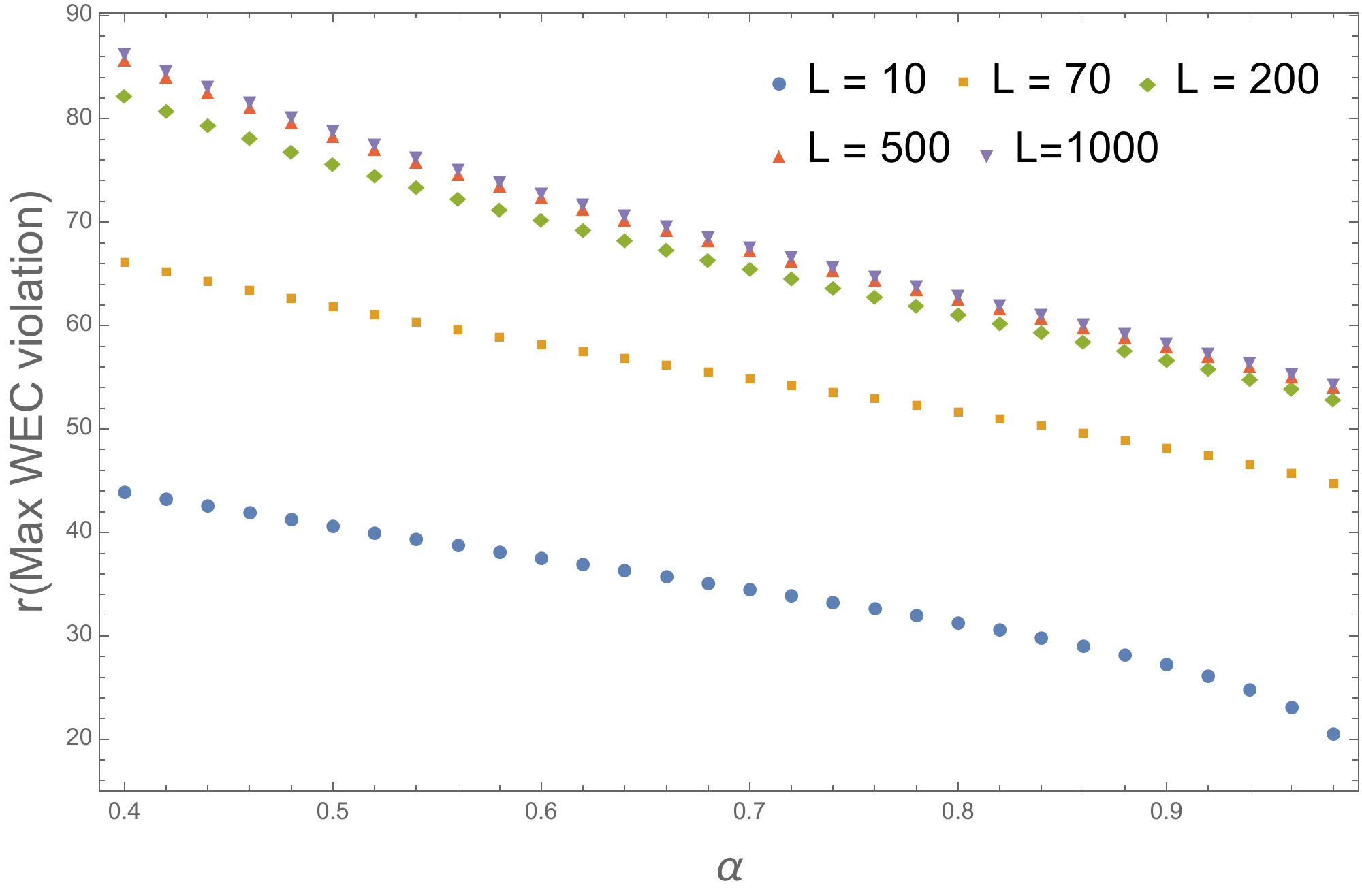}
\caption{The position of the maximum of weak energy condition violation, as a function of $\alpha$, for various values of $L$. }
\label{fig:r_WECviolation}
\end{figure}

Because the origin of the peak in curvature comes from the effective energy-momentum tensor, it is useful to investigate what happens to the weak energy condition that, as we recall, was satisfied by the Hayward metric.
As it turns out, the modification (\ref{eq:G}) introduces a violation of the weak energy condition in a small region confined around the inner horizon. As shown in Fig.~\ref{fig:energy_violation}, the sum of the density profile with the tangential pressures there fails to be non-negative. The amount by which the condition is violated increases as $L$ increases, and as $\alpha$ approaches 1.
\begin{figure}
\centering
\includegraphics[width=0.45\textwidth]{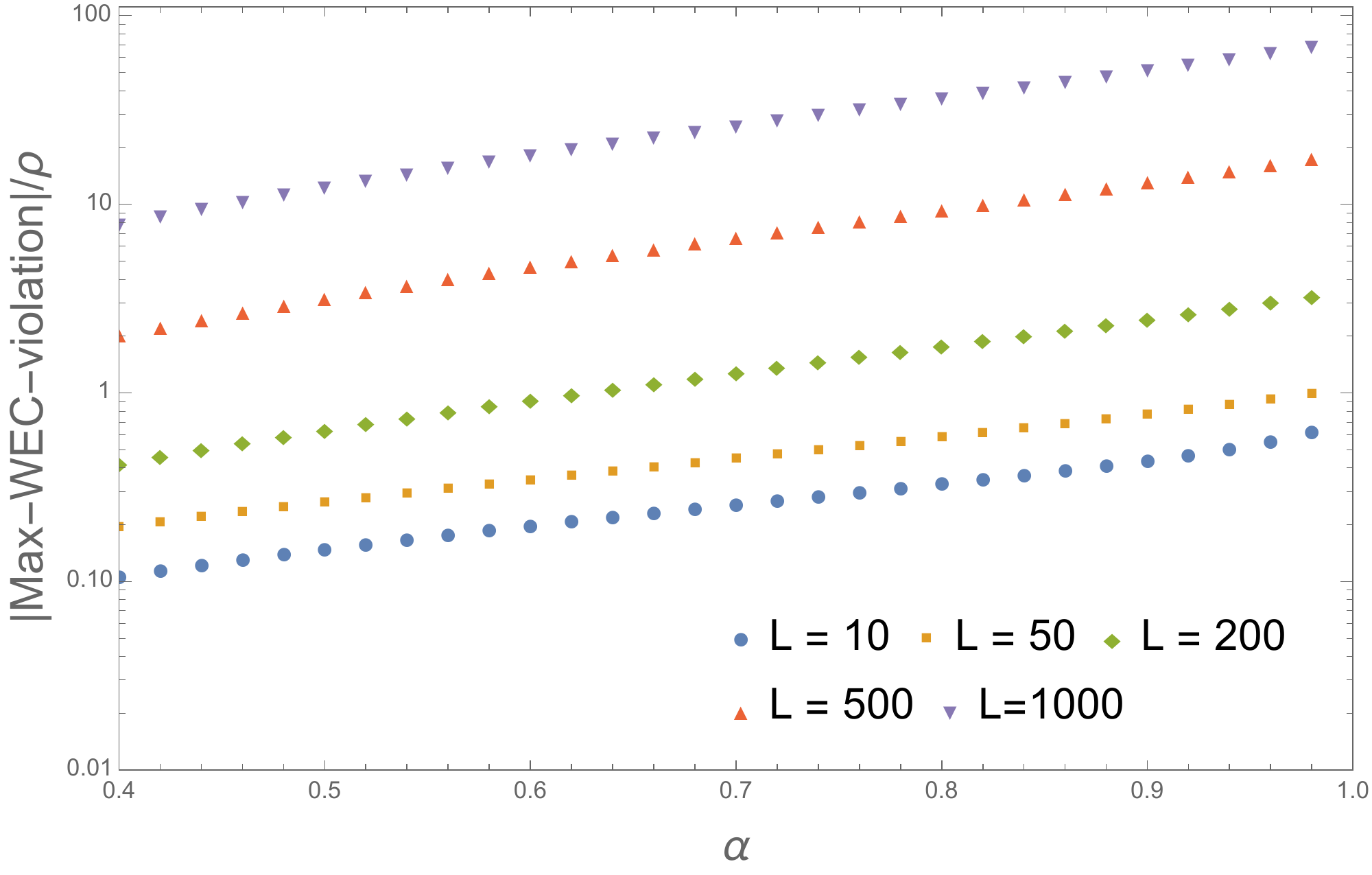}
\caption{Maximal violation of the weak energy condition, normalised by the energy density $\rho$, as a function of the $\alpha$ and for different values of $L$. Here $m=10^5$ (Planck units).}
\label{fig:WECL}
\end{figure}
The exact location of the violating region decreases as we increase both $\alpha$ and $L$, see Fig.~\ref{fig:r_WECviolation}.
Since the inner horizon is approximately located at $L$, the plot shows that the violation occurs around the inner horizon for $L=10$, and it remains well confined inside it for higher values.

While a violation of the weak energy condition is \emph{per se} not a problem in a model which is supposed to include quantum gravity effects, it is interesting to remark that the violation of the weak energy condition is a priori avoidable, 
according to a theorem by Dymnikova~\cite{dymnikova2002cosmological}, which we briefly review here, and use it to arrive at a condition on $G(r)$.
To review the derivation of the theorem, we parametrise the most general static, spherically symmetric line element as
\be
ds^2 = - e^{\mu(r)} dt^2 + e^{\nu(r)} dr^2 + r^2 d\Omega^2 \;.
\ee
From the Einstein equations,
\be
R_{\mu \nu} - \frac{1}{2} g_{\mu \nu} R = 8 \pi T_{\mu \nu},
\ee
we find the non-null components of the stress energy tensor 
\begin{subequations}
\label{eq:Tmunu}
\begin{eqnarray}
8 \pi T^0_0 = -8 \pi \rho(r) = e^{-\nu} \left( \frac{1}{r^2} - \frac{\nu'}{r} \right) - \frac{1}{r^2}\label{eq:T00}\\
8 \pi T^1_1 = 8 \pi p_r (r) = e^{-\nu} \left(\frac{1}{r^2} + \frac{\mu'}{r} \right) - \frac{1}{r^2}\\
\begin{split}
8 \pi T^2_2 &= 8 \pi T^3_3 = 8 \pi p_\perp (r)= \\
&= e^{-\nu} \left( \frac{\mu''}{2} + \frac{\mu'^2}{4} + \frac{\mu' - \nu'}{2r} - \frac{\mu' \nu'}{4} \right) \;,
\end{split}
\end{eqnarray}
\end{subequations}
where $\rho(r)$ is the energy density, while $p_r(r)$ and $p_\perp(r)$ are respectively the radial and the transversal pressures. Integration of Eq.~\eqref{eq:T00} yields 
\be
e^{-\nu} =  1 - \frac{2 M(r)}{r}
\ee
with
\be
M(r) = 4\pi \int_0^r dx \rho(x) x^2\;.
\ee
To match with our previous notations, we call
\be
m= \int_0^\infty dr \rho(r) r^2 \; 
\ee
the ADM mass of the spacetime. Let us make the following assumptions:
\begin{enumerate}
\item \emph{Dominant energy condition (DEC)};
\item \emph{Asymptotic flatness};
\item \emph{Regularity of the metric in $r=0$};
\item \emph{Finiteness of $\rho (r)$ for all  $r$};\label{hyp:rhoreg}
\item \emph{Finiteness of $m$}.
\end{enumerate}
The dominant energy condition holds if and only if $\rho(r) \geq |p_i|$ for $i=1,2,3$~\cite{hawking1973large}. As a consequence, hypothesis~\ref{hyp:rhoreg} ensures that the principal pressures are finite everywhere. It is clear that we could also require the finiteness of the pressures together with the weak energy condition, instead of asking the dominant energy condition to be satisfied.

These assumptions imply the following restrictions on the functions $\mu(r)$ and $\nu(r)$ that characterize the metric:
\begin{enumerate}
\item $\lim\limits_{r \to \infty}  \mu(r) = 0$;
\item $\nu(0) = 0$;
\item $\mu(0) \leq 0$;
\item $\mu'(r) + \nu'(r) \geq 0 \; \forall\, r$, with $\mu'(0)=\nu'(0)=0$.
\end{enumerate}

The function $A(r) \equiv \mu(r) + \nu(r)$ is monotonically increasing from a non-positive value $A(0) = \mu(0) \leq 0$ in the origin to the asymptotical value $A(\infty) = 0$. The explicit choice $A(0) = 0$ entails that $A(r)=0$ everywhere, i.e. $\mu(r) = -\nu(r)$, bringing us back to a family of metrics of the type in Eq.~\eqref{eq:dsa0=0} and then to the Hayward case.
This is the case mostly discussed in \cite{dymnikova2002cosmological}, and indeed most non-singular black hole metrics in the literature belong to this class.

On the other hand, here we suggest to rather work with $A(0)<0$, as in Eq.~\eqref{eq:linelement}, so to allow for the time delay in the center. As this behaviour is allowed by the conditions of the theorem, it means that it is possible in principle to introduce the time delay without ever violating the weak energy condition.
Let us see explicitly what restrictions this implies on the function $G(r)$ parametrising the metric as in 
\eqref{eq:linelement}. Since we are using Schwarzschild coordinates, we must distinguish two different cases: 
$r\in {\mathcal O} := [0, r_-)\cup (r_+,\infty)$, that is the inner core and exterior region, where $r$ is space-like and 
$\rho = - T^0_0, p_r = T^1_1$; and $r\in {\mathcal I} := (r_-, r_+)$, the trapping region where $\rho=-T^1_1$ and $p_r=T^0_0$. Let us assume that $G(r)>0$, so that the position of the horizons is still the same as before, determined by the zeroes of $F(r)$. Then, the condition $\rho \geq 0$ is given by
\be
\begin{split}
 1-(r\, F)' \geq 0, \quad &r\in {\cal O}\,;\\
G\big(1-(r F)' \big)- r F G' \geq 0, \quad &r\in {\cal I}\,.
\end{split}
\ee
The condition $\rho + p_r \geq 0$ imposes
\be
\begin{split}
& {F G'} \geq 0, \quad  r\in {\cal O} \;; \\
& {F G'} \leq 0, \quad  r\in {\cal I} \;.
\end{split}
\ee
Finally, the transversal part $\rho \geq p_\perp$ implies
\be\label{transv-cond}
\begin{split} 
& -r^2 F G'^2 + G^2\big(4-4F + 2r^2 F''\big) \\ 
& \quad + r G \big(3r\,F'G' + 2FG'+ 2r\,FG''\big) \geq 0, \quad r\in {\cal O} \;, \\
&-r^2 F G'^2 + G^2\big(4-4F + 2r^2 F''\big) \\
& \quad + r G \big(3r\,F'G' - 2FG'+ 2r\,FG''\big)\geq 0,  \quad r\in {\cal I}\;.
\end{split}
\ee
It can be easily checked that the function $G(r)$ proposed in Eq.~\eqref{eq:G} satisfies the first two conditions but not the last. Therefore, the violation of the weak energy condition associated with our proposal comes from the large transversal pressures.

It is natural at this point to ask whether it is possible to find a $G(r)$ satisfying the previous requirements and also \eqref{transv-cond}, thus including the time delay without violating the weak energy condition. While this is conceivably the case, we were not able to find an explicit exemple, the difficulty being matching with a smooth $G(r)$ the condition \eqref{transv-cond} on its derivatives with the local requirements $(i-iii)$, even relaxing the requirement $(iv)$ of a de Sitter behaviour near the origin, which we remark it is not demanded by the theorem above reviewed. We leave the question open for future work.

%--------------------------------------------------------
\section{Conclusions}
%--------------------------------------------------------

We have shown how it is possible to modify the non-singular black hole metric proposed by Hayward as to incorporate the 1-loop quantum corrections evaluated using the effective field theory approach to quantum gravity, as well as a non-trivial time delay between an observer at spatial infinity and one at rest in the central core, as motivated by physical requirements. We have derived an upper bound on the time delay induced by the requirement that the curvature remains sub-Planckian everywhere, so that the metric makes sense as an effective description of quantum gravity physics. The bound comes from a peak in the curvature caused by the large transversal pressures needed to generate the time delay. It is not very stringent: requiring for instance that the Kretschmann invariant is bounded by $0.1/\ell_{\rm Planck}^4$, taking 
 $m=10^5 m_{\rm Planck}$ and $L=10 \ell_{\rm Planck}$ gives a maximal time delay of $98\%$. 
The upper bound depends on both $m$ and $L$. It increases with $m$, and lightly decreases with $L$.

Introducing the time delay as we did, with a minimal modification of the metric that captures at the same time the 1-loop quantum correction at large scales, leads to a violation of the weak energy condition.
As for the peak in curvature, the violation of the weak energy condition originates from the large transversal pressures, 
and occurs well inside the quantum gravity region, where departures from classical general relativity are important. The size of the violation depends strongly on $L$, but not so much its localisation. On the other hand, we argued that it is possible to achieve the same results without violating the weak energy condition, should a compelling reason to do so present itself, and provided the explicit condition that a non-minimal modification of the metric would have to satisfy.

We hope that our modification of Hayward metric will be useful towards a more realistic description of non-singular black holes and the modelling of collapse and evaporation processes. 

%--------------------------------------------------------
\section*{Acknowledgements}
%--------------------------------------------------------
We kindly acknowledge support from the A*MIDEX project ANR-11-IDEX-0001-02, as well as the Samy Maroun Center For Space, Time and the Quantum.

%%%%%%%%%%%%%%%%%%%%

\providecommand{\href}[2]{#2}\begingroup\raggedright\endgroup

\end{document}